\documentclass[twocolumn,showpacs,prl]{revtex4}
\newcommand{\figurewidth}{\columnwidth}
\usepackage{graphicx}
\newcommand{\av}{_{\mathrm{av}}}
\newcommand{\sgn}{{\mathrm{sgn}}}

\begin{document}

\title{Single spin- and chiral-glass transition in vector spin glasses in
three-dimensions}

\author{L.~W.~Lee}
\affiliation{Department of Physics,
University of California,
Santa Cruz, California 95064}

\author{A.~P.~Young}
\affiliation{Department of Physics,
University of California,
Santa Cruz, California 95064}

\date{\today}

\begin{abstract}
Results of Monte Carlo simulations of XY and Heisenberg spin glass models in
three dimensions are presented. A finite size scaling analysis of the
correlation length of the spins and chiralities of both models
shows that
there is a single, finite-temperature transition at which both spins and
chiralities order.
\end{abstract}

\maketitle

There is now general consensus that a spin glass
transition occurs in three dimensional Ising spin glasses with short range
interactions at finite
temperature $T_{SG}$.  The most convincing work is that of Ballesteros et
al.~\cite{ballesterosetal:00} who performed a finite-size scaling analysis of
the correlation length, $\xi_L$, in samples of different sizes $L$.  Data for
the dimensionless ratio $\xi_L/L$ is found to intersect cleanly at $T=T_{SG}$,
as expected at a second order phase transition.

The situation is much less clear, however, for vector spin glasses. Early work
on XY~\cite{morris:86,jain:86} and Heisenberg~\cite{morris:86,olive:86} models
indicated a zero temperature transition, or possibly a transition at a very
low but non-zero temperature.
However, following earlier work of Villain~\cite{villain:75},
which emphasized the role of 
``chiralities'' (Ising-like
variables which describe the handedness of the non-collinear
spin structures),
Kawamura and Tanemura~\cite{kawamura:87}
proposed a 
chirality transition at
$T = T_{CG}\ (> 0)$, even though the spin glass transition
temperature, $T_{SG}$, is assumed to be
zero.  This scenario requires that spins and
chiralities decouple at long length scales. Kawamura and collaborators have
given numerical evidence for this scenario both for XY~\cite{kawamura:01} and
Heisenberg~\cite{kawamura:98,hukushima:00} models.

However, the absence of a spin glass transition in vector spin glass models
has been challenged. For the XY case, Maucourt and Grempel~\cite{maucourt:98}
and subsequently Akino and Kosterlitz~\cite{akino:02} found evidence for a
possible finite $T_{SG}$ from zero temperature domain wall calculations.
Furthermore, by studying the dynamics of the XY spin glass
in the phase representation,
Granato~\cite{granato:00} found that the ``current-voltage'' characteristics
exhibited scaling behavior which he 
interpreted as a transition in the spins as well as the chiralities.
For the Heisenberg model, Matsubara et al.~\cite{matsubara:01,endoh:01}, and
Nakamura and Endoh~\cite{nakamura:02} have
argued that the spins and chiralities order at the same low but finite
temperature.

Since the most successful approach to demonstrate a finite temperature
transition in the Ising case has been the scaling of the correlation
length~\cite{ballesterosetal:00} it seems useful to perform a similar analysis for
vector spin glass models. Furthermore, one can calculate the correlation
length for both the spins and chirality, and so perform the
\textit{same} analysis for \textit{both} types of ordering.
Here, we present results of these
calculations for the XY and Heisenberg models. For both models,
we find 
a single transition for both spins and chiralities at low but
finite temperature.

We take the standard Edwards-Anderson spin glass model
\begin{equation}
{\cal H} = -\sum_{\langle i, j \rangle} J_{ij} {\bf S}_i \cdot
{\bf S}_j,
\end{equation}
where the ${\bf S}_i$ are $n$-component
vectors of unit length at the sites of a simple
cubic lattice, and the $J_{ij}$ are nearest neighbor interactions with zero
mean and standard deviation unity. We consider both the XY model ($n$=2),
and the Heisenberg model ($n=3$).
Periodic boundary conditions are applied on
lattices with $N=L^3$ spins. 

The spin glass order parameter generalized to wave vector ${\bf k}$, 
$q^{\mu\nu}({\bf k})$, is defined to be
\begin{equation}
q^{\mu\nu}({\bf k}) = {1 \over N} \sum_i S_i^{\mu(1)} S_i^{\nu(2)}
e^{i {\bf k} \cdot {\bf R}_i},
\end{equation}
where $\mu$ and $\nu$ are spin components, and ``$(1)$'' and ``$(2)$''
denote two
identical copies of the system with the same interactions. From this we
determine the wave vector dependent
spin glass susceptibility $\chi_{SG}({\bf k})$ by
\begin{equation}
\chi_{SG}({\bf k}) = N \sum_{\mu,\nu} [\langle \left|q^{\mu\nu}({\bf
k})\right|^2 \rangle ]\av ,
\end{equation}
where $\langle \cdots \rangle$ denotes a thermal average and
$[\cdots ]\av$ denotes an average over disorder. The spin glass correlation
length
is then determined~\cite{palassini:99b,ballesterosetal:00}, from 
\begin{equation}
\xi_L = {1 \over 2 \sin (k_\mathrm{min}/2)}
\left({\chi_{SG}(0) \over \chi_{SG}({\bf k}_\mathrm{min})} - 1\right)^{1/2},
\end{equation}
where ${\bf k}_\mathrm{min} = (2\pi/L)(1, 0, 0)$.

For the XY model the chirality of a square is~\cite{kawamura:01}
\begin{equation}
\label{kappa_mu}
\kappa_i^\mu = {1 \over 2\sqrt{2}} \sum_{\langle l,m \rangle}^{\hspace{7mm}\prime}
\sgn(J_{l m}) \sin (\theta_l - \theta_m),
\end{equation}
where $\theta_l$ is the angle characterizing the direction of spin
${\bf S}_l$, and the sum is over the four bonds around the elementary plaquette
perpendicular to the $\mu$ axis and whose ``bottom left'' corner is
site $i$.
The chiral glass susceptibility is then given by
\begin{equation}
\label{chisg}
\chi_{CG}^\mu({\bf k}) =  N [\langle \left| q_{c}^\mu({\bf k})\right|^2
\rangle ]\av ,
\end{equation}
where the chiral overlap $q_{c}^\mu({\bf k})$ is given by
\begin{equation}
\label{qc}
q_{c}^\mu({\bf k}) = {1 \over N} \sum_i  \kappa_i^{\mu(1)} \kappa_i^{\mu(2)}
e^{i {\bf k} \cdot {\bf R}_i}.
\end{equation}
We define
the chiral correlation lengths $\xi^\mu_{c,L}$ by
\begin{equation}
\label{xi_c}
\xi^\mu_{c,L} = {1 \over 2 \sin (k_\mathrm{min}/2)}
\left({\chi_{CG}(0) \over \chi_{CG}^\mu({\bf k}_\mathrm{min})} - 1\right)^{1/2},
\end{equation}
in which $\chi_{CG}({\bf k}=0)$ is independent of $\mu$.
Note that $\xi^\mu_{c,L}$ will, in general, be different for $\hat{\mu}$ along
${\bf k_\mathrm{min}}$ (the $\hat{x}$ direction) and perpendicular to ${\bf k}$. We 
denote these two lengths by $\xi^\parallel_{c,L}$ and $\xi^\perp_{c, L}$
respectively.

\begin{figure}
\includegraphics[width=\figurewidth]{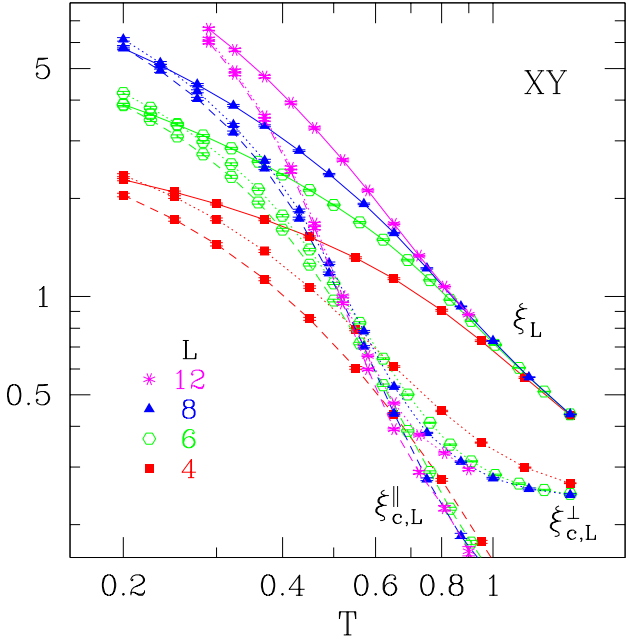}
\caption{
A plot of the spin ($\xi_L$) and chiral ($\xi^\parallel_{c,L}$ and
$\xi^\perp_{c, L}$) correlation lengths for different sizes and temperatures
for the XY spin glass. The solid lines connect data for
$\xi_L$, the dotted lines connect $\xi^\perp_{c, L}$, and the dashed lines
connect $\xi^\parallel_{c,L}$.
}
\label{xysg_xi_all}
\end{figure}

For the Heisenberg spin glass, Kawamura~\cite{kawamura:98}
defines the local chirality
in terms of three spins on a line as follows:
\begin{equation}
\kappa_i^\mu = {\bf S}_{i+\hat{\mu}} \cdot {\bf S}_i \times {\bf
S}_{i-\hat{\mu}}.
\label{eq:chiral_heis}
\end{equation}
The chiral glass susceptibilities and correlation lengths are then given in
terms of the $\kappa_i^\mu$ by
Eqs.~(\ref{chisg})-(\ref{xi_c}), as for the XY model.

\begin{figure}
\includegraphics[width=\figurewidth]{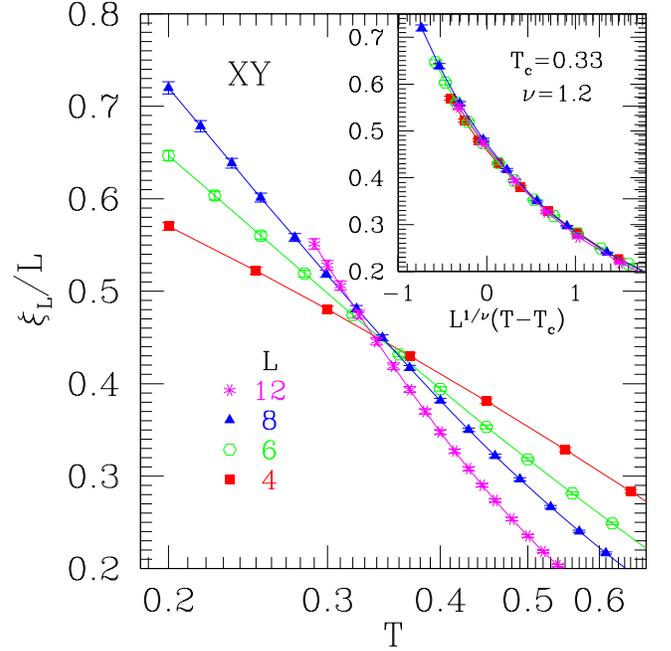}
\caption{
Plot of the spin glass correlation length $\xi_L$ divided by $L$ for the XY
spin glass. The data intersects at $T \simeq 0.34$, implying that there is a
spin glass transition at this temperature. The inset shows a scaling plot
according to Eq.~(\ref{eq:fss}) with $T_{SG} = 0.33$ and $\nu = 1.2$.
}
\label{xysg_xi_L}
\end{figure}

We use parallel tempering~\cite{hukushima:96,marinari:98b}
Monte Carlo to go down to the low
temperatures that are needed,
and study sizes from $L=4$ to 12.
To test for
equilibration~\cite{katzgraber:01} we require that the following
relation~\cite{katzgraber:01c},
\begin{equation}
[ q_l - q_s ]\av
=  {2\over z} T\, [U]\av,
\end{equation}
valid for a Gaussian bond distribution, is satisfied. Here $U$ is the 
energy per spin,
$q_l = (1/N_b)\sum_{\langle i, j \rangle}\langle
{\bf S}_i \cdot {\bf S}_j \rangle^2$
is the ``link overlap'',
$q_s = (1/N_b)\sum_{\langle i, j \rangle}\langle
({\bf S}_i \cdot {\bf S}_j)^2 \rangle$
where $N_b
= (z/2)N$ is the number of nearest neighbor bonds, and $z\ (=6\ \mbox{here})$
is the
lattice coordination number. We averaged over 1000 samples, except for 
the following cases: XY, $L=12$, 601 samples,
Heisenberg: $L=8$, 436 samples, and $L=12$, 331
samples.
The number of sweeps that each set of spins performed varied from
6000 for the small sizes to 300000 for $L=12$.

Since $\xi_L/L$ is dimensionless it has the finite size scaling form
\begin{equation}
{\xi_L \over L} = \widetilde{X}\left(L^{1/\nu}(T - T_{SG})\right) ,
\label{eq:fss}
\end{equation}
where $\nu$ is the correlation length exponent.
Note that there is
no power of $L$ multiplying the scaling function $\widetilde{X}$, as there
would be for a quantity with dimensions.
There are
analogous expressions for the chiral correlation lengths.
From Eq.~(\ref{eq:fss})
it follows that the data for $\xi_L/L$ different sizes come
together at $T=T_{SG}$. In addition, they
are also expected to splay out again on the
low-$T$ side if there is spin glass order below $T_{SG}$.

Next we discuss the results, starting with the XY spin glass. Data for the
various correlation lengths are shown in Fig.~\ref{xysg_xi_all}. One sees
that the chiral correlation lengths are smaller than the spin glass
correlation length at the higher temperatures, but increase faster on lowering
$T$, such that, for a given $L$,
all the lengths become comparable at the lowest temperatures
simulated. Furthermore the two chiral correlation lengths (parallel and
perpendicular to ${\bf k}$) become indistinguishable at lower $T$ and larger
sizes, as one would expect.

The data for $\xi_L/L$, shown in Fig.~\ref{xysg_xi_L},
intersects at a well defined temperature $\simeq
0.34$ and splay out at lower temperatures, which, according to
Eq.~(\ref{eq:fss}), implies a transition at this temperature. We
find
\begin{equation}
T_{SG} = 0.34\, \pm\, 0.02 \qquad \mbox{XY spin glass} .
\end{equation}
Note the $L=12$ data intersects at somewhat lower $T$, implying that
corrections to finite size scaling may still be significant for this range of
sizes.
Fig.~\ref{xysg_xi_L} provides compelling evidence, in our view, that there is
finite spin glass transition temperature in a three-dimensional XY spin glass,
in contrast to the claim in most of the literature. The inset 
to Fig.~\ref{xysg_xi_L} shows the data collapses well according to
Eq.~(\ref{eq:fss}) with $\nu = 1.2 \pm 0.2$. In this paper, error bars do not
include systematic effects which are hard to estimate. Given the lower
intersection point of the $L=12$ data, it is possible that $T_c$ could be lower
than that estimated here, in which case the value of $\nu$ would be increased
perhaps to the Ising value\cite{ballesterosetal:00} $2.15 \pm 0.15$.

\begin{figure}
\includegraphics[width=\figurewidth]{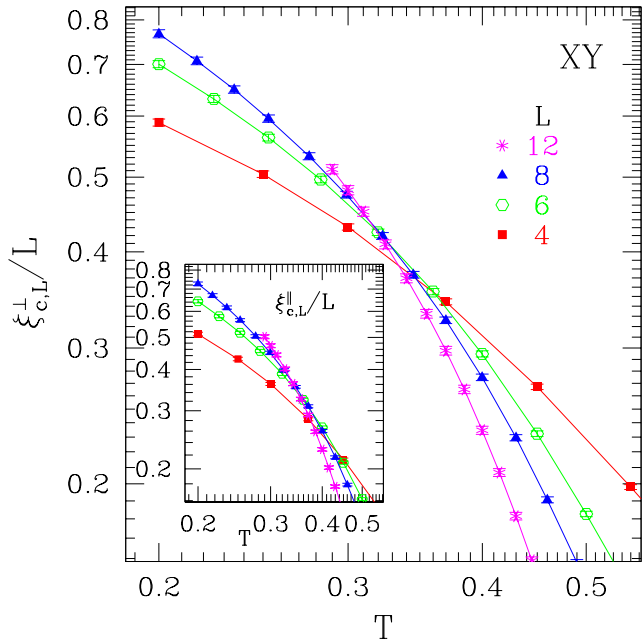}
\caption{
The main figure shows data for the perpendicular chiral correlation length
$\xi^\perp_{c,L}$ divided by $L$, for sizes $4 \le L \le 12$ for the XY spin
glass.
There are intersections at about the same
temperature as that found for the spins in Fig.~\ref{xysg_xi_L}, but evidently
with some corrections to scaling. The inset shows
analogous data for the parallel chiral correlation length,
$\xi^\parallel_{c,L}$.
}
\label{xysg_xi_chiralyz_L}
\end{figure}

Data for the chiral correlation lengths for the XY model is shown in
Fig.~\ref{xysg_xi_chiralyz_L}. Though the intersections are not quite as clean,
they occur at about the same
temperature as found above for the spin glass correlation length. For both
spin and chiral correlations,
they occur at a slightly lower $T$ for the $L=12$ data.
Altogether, the evidence is good that there is a finite chiral glass
transition at (or very close to) $T_{SG}$. Collapsing the data we find
the chiral correlation length exponent is
$\nu_c = 1.3 \pm 0.3$, which is compatible with our estimate for the spin
correlation exponent $\nu$.
Note that if the spins order then
the chiralities \textit{must} also order, assuming a non-collinear state, and
so 
$T_{CG} \ge T_{SG}$.

We are not aware of any estimates of transition temperatures for the XY spin
glass with Gaussian couplings, though for the $\pm J$ model Kawamura and
Li~\cite{kawamura:01} find $T_{CG} = 0.39 \pm 0.03$, somewhat higher than
ours. Since, for the Ising spin glass,
$T_c$ is somewhat higher for the $\pm J$ model than the Gaussian model,
our estimate of $T_c$ is probably compatible with Kawamura and Li's.
However, we emphasize that, in contrast to them, we find simultaneous
ordering of the spins and chiralities.

\begin{figure}
\includegraphics[width=\figurewidth]{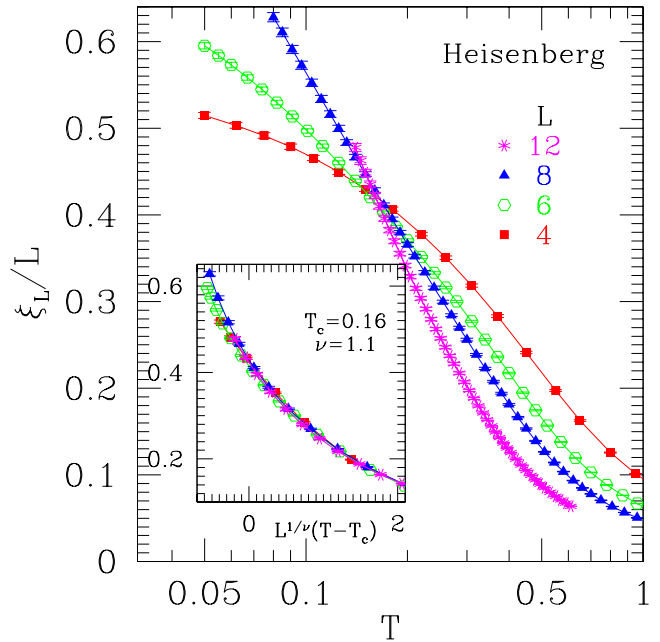}
\caption{
Data for the spin glass correlation length $\xi_L$, divided by $L$ for the
Heisenberg spin glass. The intersections imply that $T_{SG} \simeq 0.16$.
The inset shows a scaling plot according to Eq.~(\ref{eq:fss})
with $T_{SG} = 0.16$ and $\nu = 1.1$.
}
\label{hsg_xi_L}
\end{figure}

Next, we go on to our results for the Heisenberg spin glass.
As for the XY model,
the spin glass correlation length is larger at higher temperatures but the
chiral correlation length grows faster and is comparable to the spin
correlation length at the lowest temperatures.
Figure~\ref{hsg_xi_L} shows data for $\xi_L/L$,  which intersect at a common
temperature indicating a 
finite spin glass transition temperature $T_{SG}$ which we estimate to be
\begin{equation}
T_{SG} = 0.16\, \pm \, 0.02 \qquad \mbox{Heisenberg spin glass} .
\end{equation}
The inset 
to Fig.~\ref{hsg_xi_L} shows the data collapses quite well according to
Eq.~(\ref{eq:fss}) with $\nu = 1.1 \pm 0.2$.

\begin{figure}
\includegraphics[width=\figurewidth]{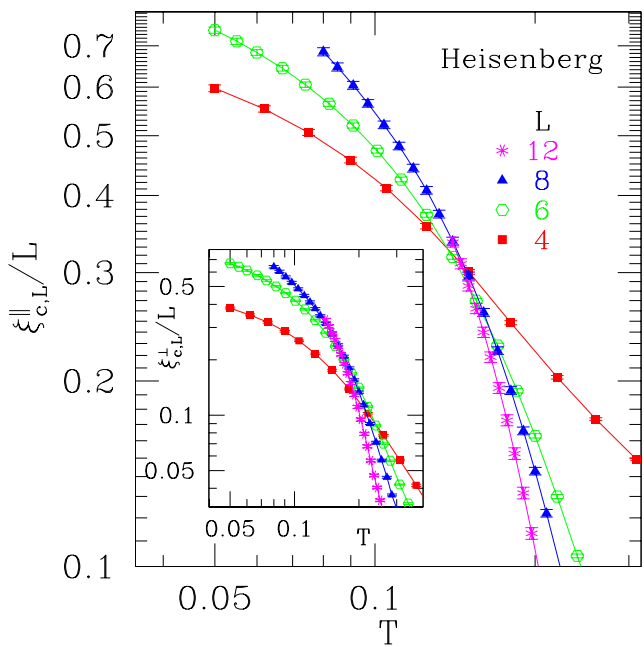}
\caption{
The main figure shows data for the parallel chiral correlation length
$\xi^\perp_{c,L}$ divided by $L$ for the Heisenberg spin glass.
There is an intersection at $T \simeq 0.14$, close to that found for the spin
glass correlation length $\xi_L$
in Fig.~\ref{hsg_xi_L}.
The inset shows
analogous data for the perpendicular chiral correlation length,
The data intersect but not as cleanly as for $\xi^\perp_{c,L}$ or 
$\xi_L$.
}
\label{hsg_xi_chiralx_L}
\end{figure}

Figure.~\ref{hsg_xi_chiralx_L} shows that the data for the chiral correlation
lengths indicate a transition at about the same value, with the intersections
being cleaner for the parallel than for the perpendicular correlation length.
Our estimate for the chiral correlation length exponent is $\nu_c = 1.3 \pm
0.3$. As for the XY spin glass, $T_c$ may be somewhat lower than that found
here, which would lead to a larger value of $\nu$.

Our value for the transition temperature agrees well with values of
$T_{CG}$ given by Kawamura~\cite{kawamura:98}, $ 0.157 \pm 0.01$, and Hukushima and
Kawamura~\cite{hukushima:00}, $0.160 \pm 0.005$, though, unlike those authors,
we claim that the spins, as well as the chiralities, order at this
temperature. For the $\pm J$ model, Endoh et al.~\cite{endoh:01} find a
spin glass transition at $T_{SG} = 0.19 \pm 0.02$ while Nakamura and
Endoh~\cite{nakamura:02} find both chiral and spin glass ordering for
$T \simeq 0.21$.

To conclude, by analyzing data for the spin and chiral correlation lengths,
we have argued
that there is a single phase transition, at
which both spins and chiralities order, in the XY and Heisenberg spin
glasses in three dimensions. 
In our view, the evidence for a spin glass transition is at
least as strong as that for a chiral glass transition.
Spin--chirality decoupling does not seem to occur.
The present work used quite modest work
station facilities, so it would be feasible to extend these results to
larger sizes by a major computational effort.

Why has this simple picture of a single finite temperature transition in
vector spin glasses in three dimensions not been generally accepted before?
One reason is that $T_{SG}$ is very low compared with the mean field value
$T_{SG}^{MF}$ ($\simeq 1.22$ for XY
and $0.82$ for Heisenberg). Until the advent of
parallel tempering~\cite{hukushima:96}
it was difficult to reach the actual $T_{SG}$ in
simulations. Furthermore, the commonly used Binder ratio
does not seem to be very useful~\cite{shirakura:02}
for vector spins, and $T=0$ domain
wall calculations are plagued by uncertainties over the optimal choice of
boundary conditions~\cite{akino:02}. We argue that, as for the Ising spin
glass~\cite{ballesterosetal:00}, finite size scaling of the correlation
length is the
optimal technique to use.

\begin{acknowledgments}
We would like to thank Helmut Katzgraber for helpful discussions, and both 
he
and H.~Kawamura for comments on a preliminary version of the manuscript. We
acknowledge support
from the National Science Foundation under grant DMR 0086287.
\end{acknowledgments}

\bibliography{refs}

\end{document}